\documentclass[superscriptaddress,amsmath, amssymb, amsfonts, twocolumn, footinbib]{revtex4-1}
\usepackage[german,american,english]{babel}
\usepackage{graphicx}
\usepackage{graphics}
\usepackage{dcolumn}
\usepackage{multirow}
\usepackage{bm}
\usepackage{latexsym}
\usepackage{amssymb}
\usepackage{amsmath}
\usepackage{amsfonts}
\usepackage{layout}
\usepackage{verbatim}
\usepackage{epsfig}
\usepackage{graphicx}
\usepackage{amsbsy}
\usepackage{lipsum}
\usepackage[toc]{appendix}
\usepackage{xcolor}

\newcommand{\bea}{\begin{eqnarray*}}
	\newcommand{\eea}{\end{eqnarray*}}
\newcommand{\bne}{\begin{equation*}}
\newcommand{\ede}{\end{equation*}}
\newcommand{\ba}{\arraycolsep 0.3ex \begin{array}{rl}}
\newcommand{\ea}{\end{array}}
\newcommand{\dps}{\displaystyle}
\newcommand{\bnen}{\begin{equation}}
\newcommand{\eden}{\end{equation}}
\newcommand{\bean}{\begin{eqnarray}}
\newcommand{\eean}{\end{eqnarray}}
\newcommand{\bsen}{\begin{subequations}}
	\newcommand{\esen}{\end{subequations}}

\newcommand{\bna}{\begin{array}}
	\newcommand{\eda}{\end{array}}
\newcommand{\bnm}{\begin{enumerate}}
	\newcommand{\edm}{\end{enumerate}}

\newcommand {\bkt} [1] {\langle #1 \rangle}

\newcommand {\dbkt} [2] {\langle #1 | #2 \rangle}

\newcommand {\pd} [2] {\frac{\partial #1}{\partial #2}}

\usepackage[utf8]{inputenc}

\begin{document}

\title{Topological nature of the proper spin current and the spin-Hall torque}
\author{Hong Liu}
\affiliation{School of Physics and Australian Research Council Centre of Excellence in Low-Energy Electronics Technologies, UNSW Node, The University of New South Wales, Sydney 2052, Australia}
\author{James H. Cullen}
\affiliation{School of Physics, The University of New South Wales, Sydney 2052, Australia}
\author{Dimitrie Culcer}
\affiliation{School of Physics and Australian Research Council Centre of Excellence in Low-Energy Electronics Technologies, UNSW Node, The University of New South Wales, Sydney 2052, Australia}
\begin{abstract}
Spin currents are key to spin torque devices, but determining the proper spin current is non-trivial. Here we derive a general quantum-mechanical formula for the intrinsic proper spin current showing that it is topological and can be finite in the gap. For topological insulators with an out of plane magnetization and the chemical potential in the surface state gap, the net spin torque is determined by the competition between the topological spin-Hall torque due to the bulk and the topological Edelstein effect due to the surface states. We also discuss spin-3/2 hole quantum wells.
\end{abstract}

\date{\today}

\maketitle




\textit{Introduction}. The spin-Hall effect (SHE) has come under renewed scrutiny thanks to recent developments in spin torque devices \cite{Manchon-RMP-2019}. Following its prediction \cite{DYAKONOV1971, DYAKONOV1971459, SHE-Hirsch-PRL-1999, Shuichi-Science-2003, Sinova-Dimi-PRL-2004}, the SHE was observed in semiconductors \cite{Exp-SHE-PRL-2005,Exp-SHE-Science-2004} and metals \cite{Bi-SHE-APL,ESHE-Copper-PRL-2011,Intrinsic-SH-metal-2011,Gaint-SHE-Ta-Sci,Giant-SHE-SOT-APL-2012,Giant-SHE-CuBi-PRL-2012,Hoffmann-PRL,SHE-Metal-APL-2014,SHE-Exp-Pt-ACSnano-2022,Giant-SHE-Metal-Adv-Mat-2022}, 
the latter facilitated by the discovery of the inverse SHE \cite{ISHE-Saitoh-JAP-2007,Kimura-ISHE-PRL-2007,ISHE-Saitoh-Natcomm-2012}, and has taken off once more in the context of spin torques and magnetic random access memory \cite{Roadmap-SOT-Review,ISHE-Bi2Se3-Nano-Lett,SC-MRAM-APL,ISHE-Phys.Rev.Research-Bi2Se3-Shuji,Yang-Gao-arXIv,Spin-SC-Rev-2023,Takahashi_2008-Rev,SHE-Device-NMat,RevModPhys.87.1213-Sinova-SHE}. The spin-Hall torque due to the spin current induces magnetization dynamics in thin ferromagnetic layers employed in spintronic memory devices \cite{ST-SHE-PRL,SC-SHT,Ohno-Current-induced-Nat-Mat-2012,SOT-Review-APR} and is believed to be strong in many topological materials \cite{Strong-SOT-Nat-Com-2022,Manchon-Nat-2014,Manchon-Nat-Comm-2018,Large-SOT-Matter-2021,Large-Damping-SOT-APL-2021,Control-SOT-Nat-Phys-2017}. In particular the bulk states of topological insulators (TI) are expected to produce a sizable spin-Hall torque inducing spin dynamics at TI/ferromagnet (FM) interfaces \cite{James-SOT,Manchon-SOT-Bulk-TI-PRB-2018, NE-Spin-texture-TI-PRB-2015,Manchon-SOT-3DTI-prb-2018,SOT-2D-Tretiakov-PRB-2017}, which is where the largest spin torques have been observed to date, including room-temperature magnetization switching \cite{Manchon-Nat-Mat-2018,SOT-Switch-Nat-Mat-2018,SOT-Switch-PRL-2017,SOT-Switch-Nat-Commun-2017,QST-PRL-2021,SOT-2D-magnet-NL-2020,SOT-PhysRevResearch-2020,Nikolic-PRX-QSOT-2021,SOC-STT-PRL-2012,Nikolic-PRL-ST-2021}. 

Generating a spin current typically requires spin-orbit coupling, which causes spin precession and hence non-conservation. Whereas the conventional definition of the spin current is the product of the spin and velocity operators \cite{Toplogical-SC-PRB-Schmeltzer,Spin-Hall-inhomo-E,Mingche-PRB-SC-2006, Conventional-SC-Sci-Rep,PhysRevB.100.245430,PhysRevB.97.195127,PhysRevLett.114.107201,PhysRevLett.117.146403,PhysRevB.71.245327,PhysRevB.74.085315} or redefined velocity operator \cite{Ghosh-PRB-2021},
spin precession generates a torque \cite{Defination-SC-PRB-2005-XC,Sugimoto-CSC-PRB,Kleinert_2006,Conserved-SC-PRB-2006,Universal-SC-Sci-Rep,Defintion-SC-PRL-2006-Qian,CSC-2D-hole-Qian-PRB-2008,STC-2DEG-PRB,QSHE-SOC-PRL-2006-Shuichi,CSHE-PRB-Kagome,Torque-CSH-PRB-2009,SC-ST-SS-prb-2017}, which makes the conventional spin current physically meaningless: its relationship to spin accumulation is not obvious \cite{Spin-accumlation-SHE-Shun-Qing-PRB-2004, SH-accumulation-Sinova-PRL-2005, Kleinert-2006,Khaetskii_PRB_2014}, it does not satisfy an equation of continuity or an Onsager relation, and is nonzero even in thermodynamic equilibrium \cite{Rashba-Spin-current-PRB, Defintion-SC-PRL-2006-Qian, Akzyanov_2021-SHE-TI-bulk}. These complications can be avoided by calculating the spin response directly without resorting to the spin current \cite{Tatara-PRB-2018,Tatara-PRB-Letter,Spin-toroidization-XiaoDi-PRB-2018,Spin-M-Quadrupole-prb-Yanase-PRB-2019}, which is appropriate when the quantity of interest is the spin accumulation. However, in systems experiencing a spin-Hall torque there is no spin accumulation, and physical insight can only be obtained by determining the spin current. This is particularly true for TI/FM devices, where the chemical potential lies in the bulk TI conduction band \cite{PhysRevB.94.014435,Band-TI/FM-Nano-Lett} and bulk transport dominates in a certain parameter regime \cite{Bulk-Tran-Dominate-Sci-Rep}. Although the spin-Hall torque due to the bulk is believed to be strong \cite{ISHE-Bi2Se3-Nano-Lett} this has never been calculated and proven. In this context a calculation of spin currents is indispensable in interpreting spin-torque experiments: it can reveal whether they are zero or finite, whether they change sign under certain circumstances, as well as their variation in different materials and with system parameters. To this end one needs to evaluate the proper spin current, which takes into account the torque dipole arising from spin precession \cite{Dimi-PRL-2004, SC-Shuichi-PRB-2004, Anisotropic-SHE-PRL-2010, PhysRevLett.109.246604, Conserved-SC-Mott-PRB-Cong-2018, Defintion-SC-PRL-2006-Qian,CSC-2D-hole-Qian-PRB-2008,SHE-insulator-PRB-2020,Cong-CC-PRB}. Its calculation is subtle, involving matrix elements of the position operator between Bloch states, and to date no quantum mechanical blueprint exists for calculating the proper spin current.

In this work we seek to remedy this shortcoming by: (i) developing a \textit{fully quantum mechanical} blueprint for calculating the intrinsic proper spin current (IPSC) using the definition of Ref.~\cite{PhysRevLett.96.076604}, yielding a formula that is ideal for band structure calculations; (ii) determining the size and structure of the spin current and its implications for the spin-Hall torque in topological materials. We find that part of the torque dipole and conventional spin current cancel out. Only a topological contribution remains, which comes in equal parts from the conventional spin current and the torque dipole. Our central result is the general expression for the IPSC
\begin{equation}\label{CS-main}
    \boldsymbol{\mathcal{J}}^l =\frac{e \bm E}{\hbar} \times \sum_{m{\bm k}} f_{m{\bm k}} \bm\Sigma^{l}_{m{\bm k}},
\end{equation}
where ${\bm E}$ is the external electric field, $m$ the band index, $f_{m{\bm k}} \equiv f(\varepsilon_{{m\bm k}})$ the equilibrium Fermi-Dirac distribution, $\varepsilon_{m{\bm k}}$ the band dispersion, and ${\bm k}$ the wave vector. Taking ${\bm E} \parallel {\bm x}$ and the transverse spin current to be along ${\bm y}$ with ${\bm z}$ direction spin polarization, we can express $\bm\Sigma^{l}_{m{\bm k}}$ as
\begin{equation}
 \Sigma^{z,z}_{m{\bm k}} = \mathrm{i} \sum_{n \ne m} s^z_{nn}(\mathcal{R}^x_{mn}\mathcal{R}^y_{nm}-\mathcal{R}^y_{mn}\mathcal{R}^x_{nm})
\end{equation}
with the Berry connection ${\bf \mathcal R}^{\bm k}_{mn}=\langle u_{m{\bm k}}|\mathrm{i}\pd{u_{n{\bm k}}}{\bm k}\rangle$. This is the \textit{only} intrinsic contribution to the proper spin current, which is shown to flow perpendicular to the applied electric field. Since only the band off-diagonal matrix elements of the Berry connection enter $\bm\Sigma^{l}_{m{\bm k}}$ the proper spin current is gauge invariant. It can also be finite in the insulating gap, with important implications for TI/FM devices, as we show below. 



\textit{Methodology}. The system is described by a single-particle density operator $\hat{\rho}$, which obeys the quantum Liouville equation
\begin{equation}\label{QLE}
\pd{\hat{\rho}}{t}+\frac{\mathrm{i}}{\hbar}[\hat{H},\hat{\rho}]=0,
\end{equation}
where $\hat{H}$ is the total Hamiltonian of the system. We will consider an arbitrary Hamiltonian and focus on a clean system, deferring the treatment of disorder to a future publication. The conserved spin current operator $\hat{\mathcal{J}}^i_j = d/dt \, (\hat{r}_j\hat{s}^i)$ is the time derivative of the spin dipole operator. Taking trace with the single particle density matrix operator $\hat{\rho}$, we separate the conserved spin current into the conventional spin current and torque dipole contributions $ \mathcal{J}^i_j = \frac{1}{2} \, {\rm Tr} \, \hat{\rho} \, \{ \hat{s}^i, \hat{v}_j \} + \frac{1}{2} \, {\rm Tr} \, \hat{\rho} \, \{ \hat{t}^i, \hat{r}_j \}$, with the velocity operator $\hat{v}_j=d\hat{r}_j/dt$ and the torque $\hat{t}^i = d\hat{s}^i/dt$, both diagonal in wave vector in the crystal momentum representation. The conventional spin current $J^i_j =\frac{1}{2}{\rm Tr} \hat{s}^i \{ \hat{v}_j, \hat{\rho} \}$ is straightforward to evaluate. In contrast $I^i_j =\frac{1}{2} {\rm Tr} \hat{\rho} \{ \hat{t}^i, \hat{r}_j\}$ stemming from the torque requires some work to deal with the position operator $\hat{r}_j$, whose matrix elements in the crystal momentum representation couple wave vectors that are infinitesimally spaced. Substituting the matrix elements of the position operator in the Bloch representation and performing some manipulations outlined in the Supplement, we find
\begin{equation}\label{Ifin-main}
	 I^i_{j} = \mathrm{i} \, {\rm tr} \sum_{\bm k} \, t^i_{\bm k}\bigg(\pd{\rho_{{\bm k}_+ {\bm k}_-}}{Q_j}\bigg)_{{\bm Q} \rightarrow 0}.
\end{equation}
Here ${\rm tr}$ represents the sum over bands, and ${\bm k}_\pm = {\bm k} \pm {\bm Q}/2$. Note that the ordinary derivative with respect to ${\bm Q}$ appears, rather than the covariant derivative, since ${\bm Q}$ represents the infinitesimal difference between two Bloch wave vectors. It is easy to prove Eq.~(\ref{Ifin-main}) is gauge invariant. 

The main challenge is determining the density matrix elements separated by the infinitesimal wave vector ${\bm Q}$. To this end we need solve Eq.~(\ref{QLE}). $\hat{H}=\hat{H}_0+\hat{H}_E$ is decomposed into band Hamiltonian $\hat{H}_0$ and electrostatic potential $\hat{H}_E$, and we work up to first order in the electric field. The method closely follows Refs.~\cite{Interband-coherence-PRB-2017, Rhonald-PR-Res-2022}, except we allow the density matrix to have terms off-diagonal in wave vector, and then expanding infinitesimal wave vector ${\bm Q}$ along the lines of Ref.~\cite{CIST-Dimi-PRB-2009}. By expanding in ${\bm Q}$ we obtain $H_{0{\bm k}_\pm} \approx H_{0{\bm k}} \pm \frac{{\bm Q}}{2} \cdot \frac{DH_{0{\bm k}}}{D{\bm k}}$, where the covariant derivative $\frac{DX}{D{\bm k}} = \pd{X}{{\bm k}} - \mathrm{i} [\bm{\mathcal R}, X]$ accounts for the wave-vector dependence of the basis functions in the crystal momentum representation. The quantum Liouville equation becomes
\begin{equation}\label{QKE}
\pd{\rho_{{\bm k}_+{\bm k}_-}}{t}+\frac{\mathrm{i}}{\hbar}[H_{0{\bm k}}, \rho_{{\bm k}_+{\bm k}_-}] + \frac{\mathrm{i}{\bm Q}}{2\hbar}\cdot\bigg\{\frac{DH_{0{\bm k}}}{D{\bm k}}, \rho_{{\bm k}_+{\bm k}_-} \bigg\}=d_{E{\bm k}}.
\end{equation}
The electric-field driving term, derived in Refs.~\cite{Interband-coherence-PRB-2017, Rhonald-PR-Res-2022}, is diagonal in ${\bm k}$: $d_{E{\bm k}} = (e{\bm E}/\hbar)\cdot(D\rho_{0{\bm k}}/D{\bm k})$. The equilibrium density matrix $\rho_{0{\bm k}, mn} = f(\varepsilon_{{\bm km}}) \, \delta_{mn}$.

Henceforth the strategy is straightforward. We first solve Eq.~(\ref{QKE}) for ${\bm Q} = 0$, whose solution to first order in ${\bm E}$, denoted by $\rho_{E{\bm k}}$, is known from Ref.~\cite{Interband-coherence-PRB-2017}. In the intrinsic regime this is 
\begin{equation}\label{Zero-Q-DM}
	\rho_{E{\bm k}, mn} = \frac{e{\bm E}\cdot{\bf \mathcal R}^{\bm k}_{mn} [f(\varepsilon_{m{\bm k}}) - f(\varepsilon_{n{\bm k}})]}{\varepsilon_{m{\bm k}} - \varepsilon_{n{\bm k}}}.
\end{equation}
This is enough to give us the conventional spin current $\langle J^i_j \rangle =\frac{1}{2}{\rm Tr} \hat{s}^i \{ \hat{v}_j, \hat{\rho}_E \}$. To obtain the torque dipole we need to find the density matrix to first order in ${\bm Q}$, which we denote by $\rho_{E{\bm Q}}$, and we find by solving Eq.~(\ref{QKE}) once more with the ${\bm Q}$-dependent term as the driving term
\begin{equation}\label{rhoQ}
\pd{\rho_{E{\bm Q}}}{t} + \frac{\mathrm{i}}{\hbar}[H_{0{\bm k}}, \rho_{E{\bm Q}}] = - \frac{\mathrm{i}{\bm Q}}{2\hbar}\cdot\bigg\{\frac{DH_{0{\bm k}}}{D{\bm k}}, \rho_{E{\bm k}} \bigg\}.
\end{equation}
This is solved in exactly the same way to yield 
\begin{equation}
	\rho_{E{\bm Q}} = \frac{\mathrm{i}{\bm Q}}{2}\cdot \frac{\{  [\bm{\mathcal R}_{\bm k}, H_{0{\bm k}}], \rho_{E{\bm k}} \}_{mn} }{\varepsilon_{m{\bm k}} - \varepsilon_{n{\bm k}}},
\end{equation}
 which finally gives the expression for the torque dipole as $\langle I^i_{j} \rangle = \mathrm{i}\, {\rm Tr} \, t^i (\partial \rho_{E{\bm Q}}/\partial Q_j)_{{\bm Q} \rightarrow 0}$. The sum 
$\langle J^i_j \rangle + \langle I^i_{j} \rangle$ yields $\langle \hat{\mathcal{J}}^i_j \rangle $ as given in Eq.~(\ref{CS-main}). 

\textit{Discussion}. A number of important observations are in order concerning the result of Eq.~(\ref{CS-main}). Firstly, for the proper spin current to be strictly conserved the expectation value of the torque, ${\rm Tr} (\hat{t}\hat{\rho})$ needs to cancel so that globally there is no net spin generation in the system \cite{Dimi-PRL-2004, Defintion-SC-PRL-2006-Qian}. In other words the torque density ${\rm Tr} (\hat{t}\hat{\rho})$ vanishes but the torque dipole density ${\rm Tr} (\{\hat{t}, \hat{{\bm r}}\} \hat{\rho})$ is finite. This is true for the models we consider below.

Secondly, in agreement with Ref.~\cite{Cong-CC-PRB}, the IPSC is perpendicular to the applied electric field, so the intrinsic spin current is \textit{only} a Hall current. Importantly, for a spin-$\frac{1}{2}$ system it is easy to show that Eq.~(\ref{CS-main}) simplifies to 
\begin{equation}
\mathcal{J}^z_{y}=\frac{eE_x}{\hbar}\sum_{m{\bm k}}s^z_{mm} \Omega^z_{m{\bm k}} f_{m{\bm k}},
\end{equation}
where the Berry curvature $\Omega^z_{m{\bm k}} = \mathrm{i}(\mathcal{R}^x_{mn}\mathcal{R}^y_{nm}-\mathcal{R}^y_{mn}\mathcal{R}^x_{nm})$. This expression has the same form as the result of Ref.~\cite{Cong-CC-PRB}, derived using purely semiclassical considerations, but differs by a sign. Since it is a topological quantity it can be nonzero even in the gap of an insulator, in analogy with the quantized anomalous Hall effect, as the examples below demonstrate. We note that Eq.~(\ref{CS-main}) can also be applied to charge transport by replacing $s^z_{mm} \rightarrow -e$, which recovers the Berry curvature contribution to the anomalous Hall effect, with the correct sign. The method we have devised for the torque dipole also yields the orbital magnetic moment of Bloch electrons with the correct magnitude and sign \cite{Sundaram_PRB99}. 

Thirdly, our calculation reveals that the part of  torque dipole cancels the conventional spin current exactly, which demonstrates that the conventional spin current is physically meaningless in spin-orbit coupled systems, and the spin current that is determined in conventional calculations is simply absent. 

We note that Ref.~\cite{Tatara-PRB-Letter} determined a tentative expression for the IPSC but identified a divergent term, which was subsequently discarded. This is consistent with the focus of Ref.~\cite{Tatara-PRB-Letter} being on spin accumulation rather than on an explicit calculation of the spin current. It is well known that the spin accumulation depends crucially on boundary conditions \cite{QSHE-SOC-PRL-2006-Shuichi,Defintion-SC-PRL-2006-Qian,Dyakonov-PRL, Khaetskii_Accumulation_PRB14, SHE-insulator-PRB-2020}, and Ref.~\cite{Tatara-PRB-Letter} demonstrated quantitatively that the spin accumulation can be determined without reference to the spin current. At the same time we are able to avoid the appearance of divergences, and we stress that our calculation is indispensable in systems in which the spin current does not lead to a spin accumulation. Such systems, which include TI/FM interfaces, are in fact used to infer the presence of a spin current. Since the spin current does not couple to any measurable quantity, its detection is primarily through indirect processes, for example by measuring spin-torque driven magnetization precession \cite{SC-indirect-measure-PRL,SC-indirect-Nat-SOT,PhysRevApplied.16.054031,E-mani-SHE-PRL-2008,ST-SHE-PRL-2011}, spin-current induced second-harmonic optical effects \cite{SC-SHM-indirect,SC-indirect-SHM-Nat}, the inverse spin Hall effect \cite{ISHE-PRL-Indirect-measure,ISHE-SC-indirect-APL}, and X-ray pump-probe measurements \cite{SC-Pump_probe-PRL}.

Finally, we stress that our result applies to 2D and 3D extended systems in which the current is carried by bulk states. We do not discuss edge states in this work.

\begin{figure}[tbp!]
\begin{center}
\includegraphics[trim=0cm 0cm 0cm 0cm, clip, width=0.75\columnwidth]{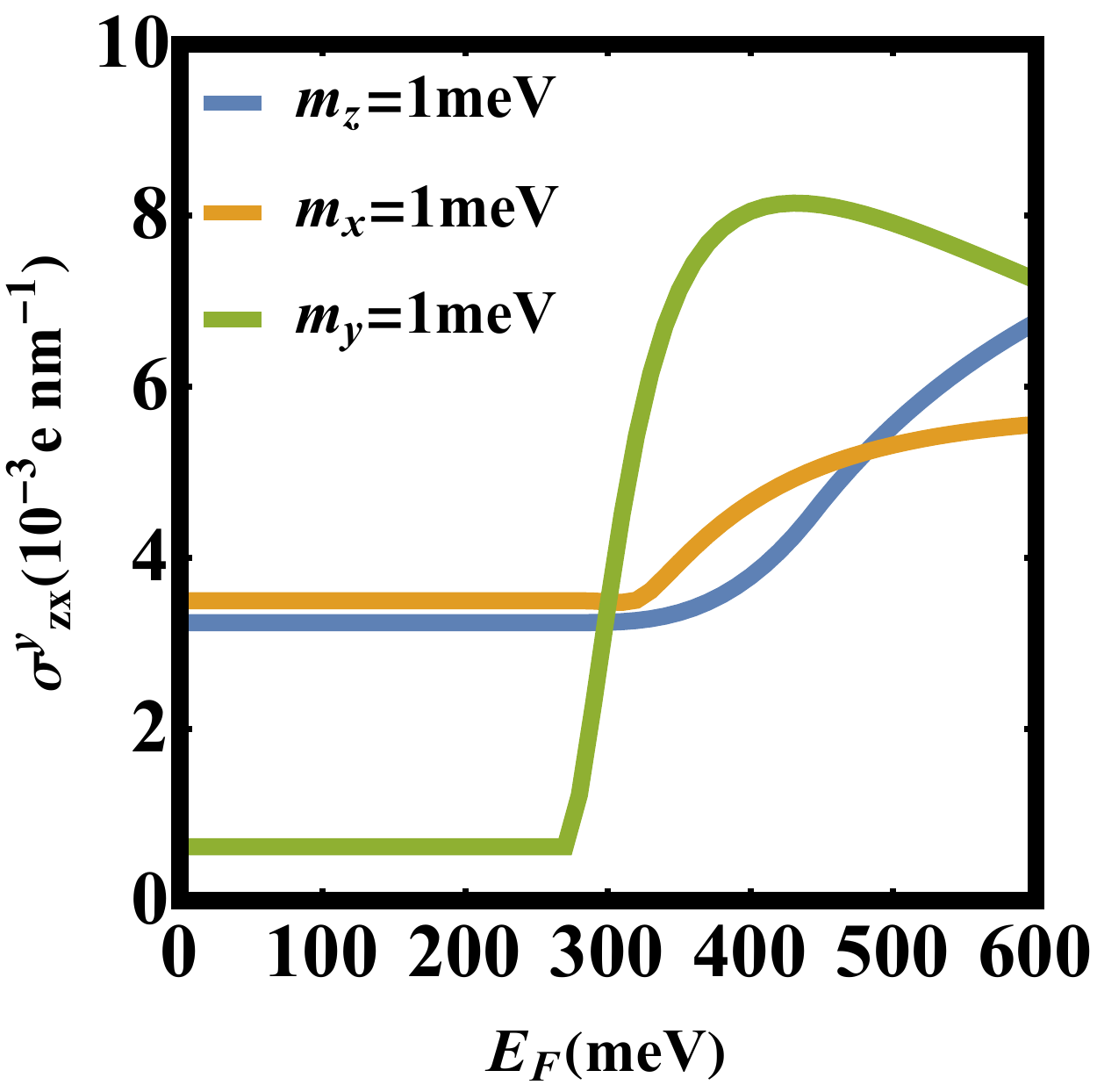}
\caption{\label{Fig1} The spin Hall conductance $\sigma^{y}_{zx}$ vs the Fermi energy $E_F$ for the TI bulk states in Bi$_2$Se$_3$ with magnetizations $\boldsymbol{m}\parallel \hat{x},\hat{y},\hat{z}$ and $|m|=1$ meV (Bi$_2$Se$_3$ parameters from Ref.~\cite{TI-bulk-Parameters}).}
\end{center}
\end{figure}

\textit{Examples: 1. Topological Insulators} We first apply our theory to the $4\times4$ topological insulator (TI) bulk Hamiltonian in Ref.~\cite{TI-bulk-Parameters} with a magnetization. $H_{0{\bm k}}=\epsilon_{\bm k}+H_\text{so}$ where $\epsilon_{\bm k}=C_0+C_1k^2_z+C_2k^2_{\parallel}$.
\begin{equation}
\ba
H_\text{so}\!=\!
\left(
\begin{array}{cccc}
-\mathcal{M} +m_z & m_-  & \mathcal{B}k_z  & \mathcal{A}k_-\\
  m_+ & -\mathcal{M}-m_z  &  \mathcal{A}k_+ & -\mathcal{B}k_z\\
\mathcal{B}k_z  & \mathcal{A}k_-  &  \mathcal{M}+m_z & m_-\\
\mathcal{A}k_+  &   -\mathcal{B}k_z & m_+  & \mathcal{M}-m_z
\end{array}
\right).
\ea
\end{equation}
The Hamiltonian is in basis $\{\frac{1}{2},-\frac{1}{2},\frac{1}{2},-\frac{1}{2}\}$. $\mathcal{M} = M_0+M_1k^2_z+M_2k^2_\parallel,  \mathcal{A}=A_0+A_2k^2_\parallel, \mathcal{B}=B_0+B_2k^2_z$. $m_z$ and $m_\pm =m_x\pm im_y$ are the magnetizations in $x,y,z$ direction. The wavevector ${\bm k}=(k\sin\theta\cos\phi, k\sin\theta\sin\phi, k\cos\theta)$ with $\theta$ the polar angle and $\phi$  azimuthal angle. We calculated the spin-current along $\hat{\bm z}$-direction normal to the plane carrying spins aligned along $\hat{\bm y}$-direction with applied electrical field along $\hat{\bm x}$-direction. Here we plot the spin Hall conductivity $\sigma^y_{zx}$ vs the Fermi energy in Fig.~\ref{Fig1}. We set the zero energy in the middle of the bulk gap, the bottom of the conduction band is at roughly $270$ meV. The Hamiltonian includes a magnetization  to remove the spin degeneracy. Though there will be a region near the interface with an induced magnetization \cite{Vobornik2011,Eremeev2013,Lang2014} we do not expect a finite magnetization in the bulk of the TI.

As shown in Fig.~1, the filled valence band states give rise to a spin current that is non zero in the band gap. Furthermore, we find that the combined spin current due to the valence and conduction band states can be additive. This differs from previous calculations that employ the conventional spin current definition which found the contributions from the valence and conduction bands to have opposite signs \cite{Manchon-SOT-Bulk-TI-PRB-2018}.

Spin currents in the bulk of TI's are of particular interest due to their application in spin torques in TI/FM systems. TI spin torques are generated through spin polarizations and currents in the TI layer: if these spins are misaligned with the magnetization of the FM layer they will apply a torque on the magnetization. Specifically, TI spin torques are believed to be driven via three primary mechanisms: The Rashba-Edelstien effect
\cite{NE-Spin-texture-TI-PRB-2015,SOT-2D-Tretiakov-PRB-2017}, the spin Hall effect
\cite{Manchon-SOT-Bulk-TI-PRB-2018} and the spin transfer torque
\cite{Kurebayashi2019,James-SOT}. The extent to which each of these effects contributes to the large spin torques recorded is yet to be settled. This is because spin torque measurements are not able to effectively determine the mechanisms via which the spin torques are generated \cite{Wang2018}.

The topological nature of IPSC results in one possibility of identifying the spin-Hall torque unambiguously. The sample can be set up with the magnetization ${\bm m} \parallel \hat{\bm z}$ perpendicular to the interface. This will open a gap in the surface state spectrum and the chemical potential can be placed in this gap - in this way it will be both in the bulk gap and the surface state gap. In this case the spin transfer torque due to the bulk will be zero \cite{James-SOT}, and the Edelstein effect due to the surface states will likewise vanish. The spin-Hall torque will have only the topological contribution, which gives rise to a field-like torque. There will be a small spin density due to the surface states, which corresponds to the quantized anomalous Hall effect and is also topological (for the surface states the charge current is equivalent to a spin density \cite{Interband-coherence-PRB-2017}), but this will give rise to a damping-like torque. The spin-Hall torque is the only field-like contribution present under these circumstances. 

When the chemical potential is in the bulk conduction band we find the spin conductivity $\sigma^y_{zx}$ of the bulk states $\sim 10^3(\hbar/2e)\Omega^{-1}\text{m}^{-1}$, one to two orders of magnitude smaller than the spin conductivities measured in experiment of $\sigma^x_{zx}=0.15-1.6\times10^5(\hbar/2e)\Omega^{-1}\text{m}^{-1}$ \cite{Manchon-Nat-2014,Manchon-Nat-Mat-2018,SOT-Switch-PRL-2017,SOT-Switch-Nat-Commun-2017} and $\sigma^y_{zx}=2\times10^5(\hbar/2e)\Omega^{-1}\text{m}^{-1}$\cite{Manchon-Nat-2014}. Furthermore, we find $\sigma^x_{zx}$ to be of a negligible magnitude compared to $\sigma^y_{zx}$. This implies that the intrinsic spin Hall effect in doped TIs should primarily give rise to a field-like torque. Since the spin conductivities calculated here are small, the intrinsic spin Hall effect contribution to the TI spin torque is expected to be negligible under most circumstances. Disorder effects have not been considered here, hence the possibility still exists that a large spin Hall effect of extrinsic origin may occur when the chemical potential is in the bulk conduction band. We defer this investigation to a later work.

\begin{figure}[tbp]
\begin{center}
\includegraphics[trim=0cm 0cm 0cm 0cm, clip, width=0.49\columnwidth]{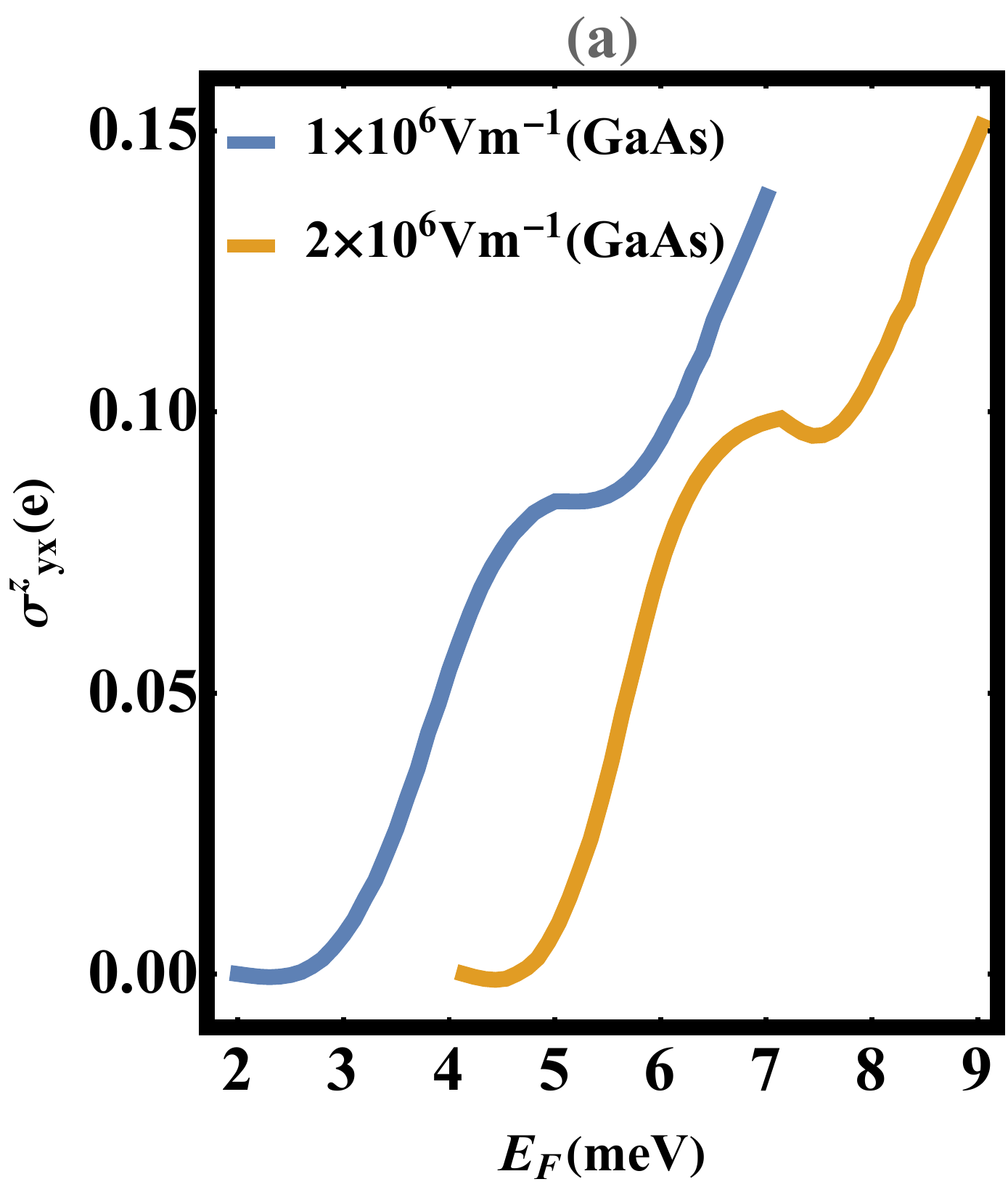}
\includegraphics[trim=0cm 0cm 0cm 0cm, clip, width=0.49\columnwidth]{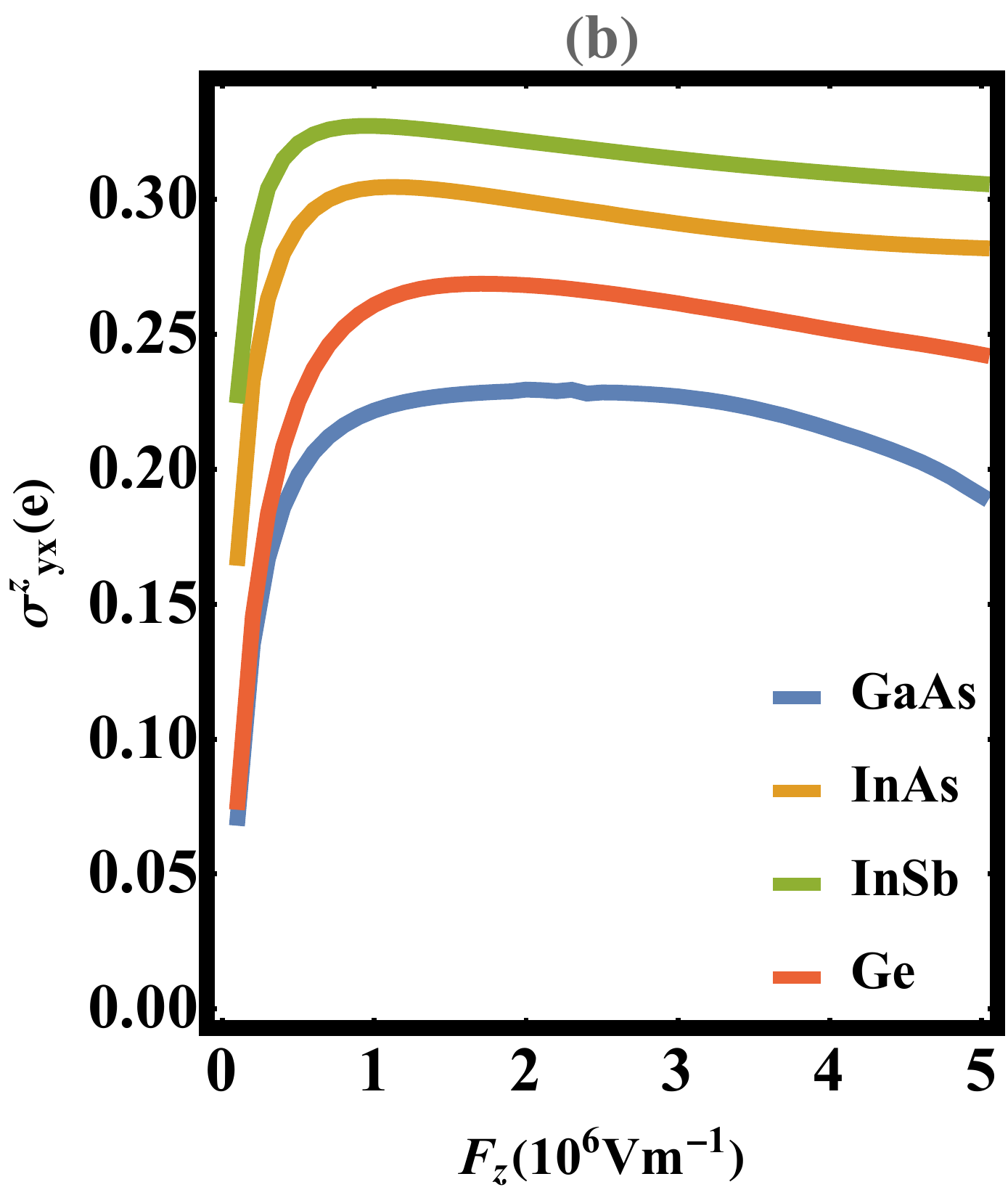}
\caption{\label{Fig2}(a) Spin Hall conductivity $\sigma^z_{yx}$ of GaAs dependence on Fermi energy with perpendicular confinement electrical field $F_z=1\times 10^6\text{V}\text{m}^{-1}$ and $F_z=2\times 10^6\text{V}\text{m}^{-1}$.
(b) Spin Hall conductivity $\sigma^z_{yx}$ for GaAs, InAs, InSb and Ge vs the top gate electrical field. $E_F$ is set to be $15,40,60,20$ meV for GaAs, InAs, InSb and Ge, respectively. All calculations in (a) and (b) with perpendicular magnetization $m_z=1$ meV and quantum well width 25 nm.}
\end{center}
\end{figure}

\textit{2. 2D hole gases}. We apply our theory to the 2D hole systems, in the presence of a constant in-plane electric field ${\bm E}=E\hat{\bm x}$ and a perpendicular magnetization $m_z=1\text{meV}$.
We start from the bulk Luttinger Hamiltonian \cite{Luttinger-1956-PR}  $H_L(k^2,k_z)$ describing holes in the uppermost valence band with an effective spin $J=3/2$. So the hole system with top and back gate in $z$-direction can be simplified as the isotropic Luttinger Hamiltonian plus a confining asymmetrical triangular potential $\hat{H}=H_L(k^2,\hat{k}_z)- e F_z z$ with $F_{z}$ the gate electric field. The Hamiltonian is expressed in the basis of $J_z$ eigenstates
$\{|+\frac{3}{2}\rangle,|-\frac{3}{2}\rangle,|+\frac{1}{2}\rangle,|-\frac{1}{2}\rangle\}$
\begin{equation}
\ba
H_L(k^2,\hat{k}_z)\!=\!
\left(
\begin{array}{cccc}
P+Q & 0  & L& M \\
0  & P+Q  & M^* & -L^*  \\  
L^* & M & P-Q &0 \\
M^* & -L & 0 & P-Q
\end{array}
\right),
\ea
\end{equation}
$ P=\frac{\hbar^2}{2m_0}\gamma_1(k^2+k^2_z), Q=-\frac{\hbar^2}{2m_0}\gamma_2(2k^2_z-k^2),L=-\sqrt{3}\frac{\hbar^2}{m_0}\gamma_3k_-k_z, M=-\frac{\sqrt{3}}{2}\frac{\hbar^2}{m_0}(\overline{\gamma}k^2_-+\delta k^2_+)$.
$\gamma_1,\gamma_2,\gamma_3$ are Luttinger parameters,
$\overline{\gamma}=\frac{\gamma_2+\gamma_3}{2}$, $\delta=\frac{\gamma_2-\gamma_3}{2}$, $k_{\pm}=k_x\pm \mathrm{i}k_y$ and $\theta=\arctan \frac{k_y}{k_x}$. We use modified infinite square well wave functions \cite{Bastard-Wavefunction} for the heavy hole (HH) and light hole (LH) states
 \begin{equation}\label{Bastard}
 	\phi_v = \frac{\sin\left[\frac{\pi}{d}\left(z + \frac{d}{2}\right)\right] \exp \left[ -\beta_v \left(\frac{z}{d} + \frac{1}{2} \right) \right]}{\pi \sqrt{\frac{e^{-\beta_v} d \sinh(\beta_v) }{2 \pi^2 \beta_v + 2 \beta_v^3}}}, 
 \end{equation}
 $v = h, l$ denote the HH and LH states and $d$ is the width of the quantum well. The eigenvalues of HH and LH are obtained by diagonalizing the matrix $ \tilde{H}$, whose elements are given as $ \tilde{H}=\langle \nu|H_L(k^2,\hat{k}_z)+V(z)|\nu'\rangle+H_m$, 
where $|\nu\rangle$ denotes the wave function Eq.~(\ref{Bastard}) and $\hat{k}_z=-i\pd{}{z}$. Fig.~\ref{Fig2}(a) shows the intrinsic spin Hall conductivity increases with increasing of the Fermi energy, where spin-polarization is along $\hat{\bm z}$. Fig.~\ref{Fig2}(b) shows $\sigma^{z}_{yx}$ dependence on the confinement electrical field, experiencing an upturn then slow down turn because  Rashba coefficient has the same trend \cite{Hong-PRL-2018}.

The spin Hall conductivity of 2D hole system has been investigated in the past \cite{CSC-2D-hole-Qian-PRB-2008,SHE-hole-Mz-PRB,Universal-SC-Sci-Rep}, with most studies taking the holes Rashba spin-orbital coupling in analogy with electrons and neglecting the HH-LH coupling. This simplistic approach is not sufficient in calculating the proper spin current. Our general formula enables us to work with the $4\times4$ Luttinger Hamiltonian and account for HH-LH coupling. We note that a method similar to ours was proposed in the context of 2D hole gases \cite{Kleinert_2006}, but the original paper erroneously obtained a zero the spin-Hall conductivity due to a failure to account for the coupling between heavy holes and light holes. In this context a recent paper \cite{Ghosh-PRB-2021} resorted to a modified spin current in order to remove the finite spin current in equilibrium and to illustrate the role of the Berry curvature in linear and non-linear spin transport.

\textit{Conclusion}. In this work, we demonstrate a new route to realizing an experimentally controllable spin-Hall effect, and apply it on the Luttinger Hamiltonian and the other well known Hamiltonians. We confirmed the topological origin the proper spin current and also provide a formula that can be easily evaluated numerically for realistic band structures. The formalism can be straightforwardly extended to the complicated case of disorder,\cite{Schwab2002, Spin-current-Polarization-2DEG-PRL-2004, Inoue_RashbaSHE_Vertex_PRB04, Khaetskii_Nonexistence_PRL06, Culcer_SJ_PRB10, PhysRevB.88.035316} and this will be done in a future publication.

\textit{Acknowledgments.} This work is supported by the Australian Research Council Centre of Excellence in Future Low-Energy Electronics Technologies, project number CE170100039. We are grateful to Qian Niu, Cong Xiao, Binghai Yan and Shuichi Murakami for enlightening discussions. 


%

\clearpage
\begin{widetext}
\begin{appendix}

\section{Simplified method for evaluating the torque dipole}
The torque dipole contribution to the spin current is given by 
\begin{equation}
\arraycolsep 0.3ex
\begin{array}{rl}
\displaystyle I^i_j = & \displaystyle \frac{1}{2} \, {\rm Tr} \, \hat{\rho} \, \{ \hat{t}^i, \hat{r}_j \} = \frac{1}{2} \, {\rm Tr} \, \{ \hat{r}_j, \hat{\rho} \} \, \hat{t}^i.
\end{array}
\end{equation}
In the crystal momentum representation $t^i$ is diagonal in wave vector. The matrix elements of the position operator in the Bloch representation are given by
\begin{equation}\label{rkk}
    {\bm r}_{{\bm k}{\bm k}'} = \mathrm{i} \, \pd{}{{\bm k}}\delta( {\bm k} - {\bm k}') + \bm{\mathcal R}_{\bm k} \delta( {\bm k} - {\bm k}'),
\end{equation}
with $\bm{\mathcal R}_{\bm k}$ the Berry connection. 

We first discuss a simplified method relying on the Pauli basis, which represents eigenstates of the spin/pseudospin operators, so the spinors are independent of wave vector. The spin/pseudospin can have an arbitrary dimensionality. Although this method has a limited applicability we focus on it due to its extreme simplicity, and the fact that it is applicable to all ${\bm k}\cdot{\bm p}$ band structure models, including the Kane model for semiconductors, the Dirac models for all forms of graphene and topological insulators, as well as Weyl semimetals, transition metal dichalcogenides, and other topological materials. The extension to an arbitrary Bloch basis is covered below. 

In the Pauli basis the Berry connection is zero, and the matrix elements of the position operator are just ${\bm r}_{{\bm k}{\bm k}'} = \mathrm{i} \, \pd{}{{\bm k}}\delta( {\bm k} - {\bm k}')$. Then:
\begin{equation}
 I^i_j = \frac{1}{2} \, {\rm Tr} \, t^i_{{\bm k}{\bm k}} \sum_{{\bm k}'}\, (r_{j, {\bm k}{\bm k}'} \rho_{{\bm k}'{\bm k}} + \rho_{{\bm k}{\bm k}'}r_{j, {\bm k}'{\bm k}})= \frac{1}{2} \, {\rm Tr} \, t^i_{{\bm k}{\bm k}} \int \frac{d^dk'}{(2\pi)^d}\, \mathrm{i}\, [\pd{}{k_j}\delta( {\bm k} - {\bm k}')] \rho_{{\bm k}'{\bm k}} + \mathrm{i} \rho_{{\bm k}{\bm k}'} \, \pd{}{k'_j}\delta( {\bm k}' - {\bm k}),
\end{equation}
where $\sum_{{\bm k}'}=\int \frac{d^dk'}{(2\pi)^d}$ with $d$ the dimension of the system. In the first term we make the replacement
\begin{equation}
\pd{}{k_j}\delta( {\bm k} - {\bm k}')] \rho_{{\bm k}'{\bm k}} \rightarrow \pd{}{k_j}[\delta( {\bm k} - {\bm k}') \rho_{{\bm k}'{\bm k}}] - \delta( {\bm k} - {\bm k}') \pd{\rho_{{\bm k}'{\bm k}}}{k_j},
\end{equation}
and in the second term we integrate the $\delta$ function by parts,
\begin{equation}
\arraycolsep 0.3ex
\begin{array}{rl}
\displaystyle I^i_j = & \displaystyle \frac{1}{2} \, {\rm Tr} \, t^i_{{\bm k}{\bm k}} \int \frac{d^dk'}{(2\pi)^d}\, \mathrm{i}\, \bigg\{ \pd{}{k_j}[\delta( {\bm k} - {\bm k}') \rho_{{\bm k}'{\bm k}}] - \delta( {\bm k} - {\bm k}') \pd{\rho_{{\bm k}'{\bm k}}}{k_j} \bigg\} - \mathrm{i} \, \pd{\rho_{{\bm k}{\bm k}'}}{k'_j} \, \delta( {\bm k}' - {\bm k}) \\ [3ex]

= & \displaystyle \frac{\mathrm{i}}{2} \, {\rm Tr} \, t^i_{{\bm k}{\bm k}} \, \bigg[ \frac{1}{(2\pi)^d} \pd{\rho_{{\bm k}{\bm k}}}{k_j} - \int \frac{d^dk'}{(2\pi)^d}\,\delta( {\bm k} - {\bm k}') \bigg( \pd{\rho_{{\bm k}'{\bm k}}}{k_j} + \pd{\rho_{{\bm k}{\bm k}'}}{k'_j} \bigg) \bigg].
\end{array}
\end{equation}
The imaginary terms must cancel, a fact that will be shown explicitly.
\begin{equation}\label{Iij}
\arraycolsep 0.3ex
\begin{array}{rl}
\displaystyle I^i_{j} = & \displaystyle \frac{\mathrm{i}}{2} \, {\rm Tr} \, t^i_{ {\bm k}{\bm k}} \, \bigg[ \frac{1}{(2\pi)^d} \pd{\rho_{{\bm k}{\bm k}}}{k_j} - \int \frac{d^dk'}{(2\pi)^d}\,\delta( {\bm k} - {\bm k}') \bigg(\pd{\rho_{{\bm k}{\bm k}'}}{k'_j} + \pd{\rho_{{\bm k}'{\bm k}}}{k_j} \bigg) \bigg].
\end{array}
\end{equation}
To evaluate this term we first make a change of coordinates ${\bm k} \equiv {\bm q}_+ = {\bm q} + {\bm Q}/2$ and ${\bm k}' \equiv {\bm q}_- = {\bm q} - {\bm Q}/2$. The wave vector derivatives take the form
\begin{equation}
	\arraycolsep 0.3ex
	\begin{array}{rl}
		\displaystyle \pd{}{{\bm k}} = \frac{1}{2} \pd{}{{\bm q}} + \pd{}{{\bm Q}} ,\quad \pd{}{{\bm k}'} = \frac{1}{2} \pd{}{{\bm q}} - \pd{}{{\bm Q}}.
	\end{array}
\end{equation}
From the above, it is immediately obvious that the two terms $
	\bigg(\pd{\rho_{{\bm k}{\bm k}'}}{k'_j} + \pd{\rho_{{\bm k}'{\bm k}}}{k_j} \bigg)$ in brackets in Eq.\ (\ref{Iij}) give the same result. The sum of the $\displaystyle \pd{}{{\bm q}}$ terms cancels the first term on the RHS of Eq.\ (\ref{Iij}), and the term we are looking for is
\begin{equation}\label{Ifinq}
	\arraycolsep 0.3ex
	\begin{array}{rl}
		\displaystyle I^i_{j} = & \displaystyle \mathrm{i} \, {\rm tr} \, \int \frac{d^dq}{(2\pi)^d} \, t^i_{{\bm q}} \int \frac{d^dQ}{(2\pi)^d} \, \delta({\bm Q}) \pd{\rho_{{\bm q}_+ {\bm q}_-}}{Q_j} = \mathrm{i} \, {\rm tr} \, \int \frac{d^dq}{(2\pi)^{2d}} \, t^i_{{\bm q}} \bigg(\pd{\rho_{{\bm q}_+ {\bm q}_-}}{Q_j}\bigg)_{{\bm Q} \rightarrow 0}.
	\end{array}
\end{equation}
Here  tr is just the spin trace and the Jacobian $J({\bm k}{\bm k}'|{\bm q}{\bm Q}) = 1$. So far no approximations, and the formula we have just obtained holds in the energy eigenstate basis as well. This is because there is no covariant derivative with respect to ${\bm Q}$, we only have the ordinary derivative. Relabel the wavevector ${\bm q}\rightarrow{\bm k}$ in Eq.~(\ref{Ifinq}), we get 
\begin{equation}\label{eq:Iij}
	\arraycolsep 0.3ex
	\begin{array}{rl}
		\displaystyle I^i_{j} = \frac{\mathrm{i}}{2} \, {\rm tr} \, \int \frac{d^dk}{(2\pi)^{2d}} \, t^i_{{\bm k}}   \bigg(\pd{\rho_{{\bm k}_+ {\bm k}_-}}{Q_j} - \pd{\rho_{{\bm k}_- {\bm k}_+}}{Q_j}\bigg)_{{\bm Q} \rightarrow 0}.
	\end{array}
\end{equation}
Since in the calculation of the torque dipole the term $\mathrm{i} \, \pd{\rho_{{\bm k}_+ {\bm k}_-}}{Q_j}$ leads to a real result, in the main text we have retained only this term and removed the factor of $1/2$ in order to simplify the notation.

The main question is how to evaluate $\displaystyle\pd{\rho_{{\bm k}_+ {\bm k}_-}}{Q_j}$. We allow the density matrix to have terms diagonal and off-diagonal in wave vector, and distinguish two wave vector scales: a small one, which accounts for inhomogeneity and finite particle size, and a large scale which accounts for scattering (scattering is not considered in this work). The density matrix is separated into a part that is \textit{nearly} off-diagonal in ${\bm k}$ and the true off-diagonal part. In the Pauli basis, the commutator of the density matrix with the band Hamiltonian is
\begin{equation}\label{commutator}
\arraycolsep 0.3ex
\begin{array}{rl}
\displaystyle \frac{\mathrm{i}}{\hbar} [\hat{H}_0, \hat\rho]_{{\bm k}_+{\bm k}_-} = & \displaystyle \frac{\mathrm{i}}{\hbar} \big(H_{0{\bm k}_+} \rho_{{\bm k}_+{\bm k}_-} - \rho_{{\bm k}_+{\bm k}_-} H_{0{\bm k}_-}\big), \\ [3ex]
\displaystyle H_{0{\bm k}_\pm} \approx & \displaystyle H_{0{\bm k}} \pm \frac{{\bm Q}}{2} \cdot \pd{H_{0{\bm k}}}{{\bm k}}.
\end{array}
\end{equation}
The kinetic equation can then be written as
\begin{equation}\label{eq:kn1}
\arraycolsep 0.3ex
\begin{array}{rl}
\displaystyle \pd{\rho_{{\bm k}_+{\bm k}_-}}{t} + \frac{\mathrm{i}}{\hbar} \, [H_{0{\bm k}}, \rho_{{\bm k}_+{\bm k}_-}] + \frac{\mathrm{i}{\bm Q}}{2\hbar} \cdot \bigg\{\pd{H_{0{\bm k}}}{{\bm k}}, \rho_{{\bm k}_+{\bm k}_-} \bigg\} = & \displaystyle d_{{\bm k}_+{\bm k}_-},
\end{array}
\end{equation}
where $d_{{\bm k}_+{\bm k}_-}$ is an arbitrary driving term. We can solve the kinetic equation to first order in ${\bm Q}$, then take the limit ${\bm Q} \rightarrow 0$. Taking initially ${\bm Q} = 0$, the initial density matrix is denoted by $\rho_{\bm k} \equiv \rho_{{\bm k}{\bm k}}$ and is diagonal in wave vector, while the correction to first order in ${\bm Q}$ is denoted by $\rho_{{\bm Q}}$ (this is done in order to keep the notation sinmple -- $\rho_{{\bm Q}}$ depends on both ${\bm k}$ and ${\bm Q}$). Since the driving term due to the electric field is diagonal in ${\bm k}$ we write it as $d_{\bm k}$.
\begin{equation}\label{eq:kn2}
\arraycolsep 0.3ex
\begin{array}{rl}
\displaystyle \pd{\rho_{\bm k}}{t} + \frac{\mathrm{i}}{\hbar} \, [H_{0{\bm k}}, \rho_{\bm k}] = & \displaystyle d_{\bm k}, \\ [3ex]
\displaystyle \pd{\rho_{{\bm Q}}}{t} + \frac{\mathrm{i}}{\hbar} \, [H_{0{\bm k}}, \rho_{{\bm Q}}] = & \displaystyle - \frac{\mathrm{i}{\bm Q}}{2\hbar} \cdot \bigg\{\pd{H_{0{\bm k}}}{{\bm k}}, \rho_{\bm k} \bigg\}.
\end{array}
\end{equation}
The term $(1/\hbar) \, \partial H_{0{\bm k}}/\partial {\bm k}$ on the RHS is recognized as the velocity operator. The solution 
$\rho_{{\bm Q}}$ can be substituted into Eq.~(\ref{eq:Iij}) to determine the torque dipole. 

\section{Extension to an arbitrary Bloch basis}

In an arbitrary Bloch basis the Berry connection is nonzero. Restoring the Berry connection to the matrix elements in Eq.~(\ref{rkk}), it is straightforward to show that
\begin{equation}\label{eq:Iijgen}
	\arraycolsep 0.3ex
	\begin{array}{rl}
		\displaystyle I^i_{j} = \frac{1}{2} \, {\rm tr} \, \int \frac{d^dk}{(2\pi)^{2d}} \, t^i_{{\bm k}} \bigg[ \mathrm{i} \, \bigg(\pd{\rho_{{\bm k}_+ {\bm k}_-}}{Q_j} - \pd{\rho_{{\bm k}_- {\bm k}_+}}{Q_j}\bigg)_{{\bm Q} \rightarrow 0} + \{ \bm{\mathcal R}_{\bm k}, \rho_{\bm k} \} \bigg].
	\end{array}
\end{equation}
The band-diagonal matrix elements of the Berry connection are gauge-dependent. The first term in the square brackets in Eq.~(\ref{eq:Iijgen}) also produces gauge-dependent terms, and these cancel the gauge-dependent terms arising from $\bm{\mathcal R}_{\bm k}$, so that $I^i_j$ in Eq.~(\ref{eq:Iijgen}) is gauge-invariant. In fact, if one sets $t^i_{\bm k} \rightarrow 1$ in Eq.~(\ref{eq:Iijgen}) one obtains the expectation value of the position operator in this representation
\begin{equation}\label{eq:rexp}
	\arraycolsep 0.3ex
	\begin{array}{rl}
		\displaystyle \bkt{r_{j}} = \frac{1}{2} \, {\rm tr} \, \int \frac{d^dk}{(2\pi)^{2d}} \, \bigg[ \mathrm{i} \, \bigg(\pd{\rho_{{\bm k}_+ {\bm k}_-}}{Q_j} - \pd{\rho_{{\bm k}_- {\bm k}_+}}{Q_j}\bigg)_{{\bm Q} \rightarrow 0} + \{ \bm{\mathcal R}_{\bm k}, \rho_{\bm k} \} \bigg].
	\end{array}
\end{equation}
The differential terms represent the phase of the wave function in more conventional evaluations, while the Berry connection represents the contribution due to the change of the basis states between infinitesimally-separated wave vectors. Eq.~(\ref{eq:rexp}) makes it clear that the operator 
\begin{equation}\label{eq:cdod}
   \mathcal{D}\{\rho\}_{{\bm k}_+ {\bm k}_-} \equiv \mathrm{i} \, \bigg(\pd{\rho_{{\bm k}_+ {\bm k}_-}}{Q_j} - \pd{\rho_{{\bm k}_- {\bm k}_+}}{Q_j}\bigg)_{{\bm Q} \rightarrow 0} + \{ \bm{\mathcal R}_{\bm k}, \rho_{\bm k} \},
\end{equation}
plays the role of a covariant derivative involving the ${\bm k}$-off diagonal matrix elements of $\rho$, with the anti-commutator appearing due to the presence of the imaginary factor $\mathrm{i}$ multiplying the expression. 

Determining $\rho_{{\bm k}}$ and $\rho_{{\bm Q}}$ in an arbitrary Bloch basis is more laborious because of the requirement of gauge invariance. In an arbitrary basis the procedure outlined in Appendix A must be revised. One cannot simply evaluate $\pd{\rho_{{\bm k}_+ {\bm k}_-}}{Q_j}$ since this is gauge-dependent. Instead, the full derivative $\mathcal{D}$ introduced in Eq.~(\ref{eq:cdod}) must be applied to the quantum Liouville equation in order to determine $\rho_{{\bm k}_+{\bm k}_-}$ directly:
\begin{equation}\label{eq:kn3}
\arraycolsep 0.3ex
\begin{array}{rl}
\displaystyle \pd{\rho_{{\bm k}_+{\bm k}_-}}{t} + \frac{\mathrm{i}}{\hbar} \, [H_{0{\bm k}}, \rho_{{\bm k}_+{\bm k}_-}] = & \displaystyle 0, \\ [3ex]
\displaystyle \pd{}{t} \, \mathcal{D}\{\rho\}_{{\bm k}_+{\bm k}_-} + \frac{\mathrm{i}}{\hbar} \, \mathcal{D} \{ [H_0, \rho] \}_{{\bm k}_+{\bm k}_-} = & \displaystyle 0.
\end{array}
\end{equation}
The procedure is lengthy and is not covered in detail here. However, the net result is that all the terms involving the Berry connection cancel, and the remaining contribution to the density matrix can be found from an equation formally equivalent to Eq.~(\ref{eq:kn2})
\begin{equation}\label{eq:kn4}
\arraycolsep 0.3ex
\begin{array}{rl}
\displaystyle \pd{\rho_{{\bm Q}}}{t} + \frac{\mathrm{i}}{\hbar} \, [H_{0{\bm k}}, \rho_{{\bm Q}}] = & \displaystyle - \frac{\mathrm{i}{\bm Q}}{2\hbar} \cdot \bigg\{\frac{DH_{0{\bm k}}}{D{\bm k}}, \rho_{\bm k} \bigg\},
\end{array}
\end{equation}
in which the ordinary derivatives $\partial/\partial {\bm k}$ in Eq.~(\ref{eq:kn1}) are replaced by the regular covariant derivative $D/D{\bm k}$ defined in the main text, which acts only on ${\bm k}$-diagonal matrix elements and takes into account the curvature of the eigenspace. This is Eq.~(7) in the main text.

\section{Explicit calculation of the proper intrinsic spin current }
For this part we will switch to energy eigenstate basis and use the notation in our inter-band coherence paper \cite{Interband-coherence-PRB-2017}. 
\subsection{Conventional intrinsic spin current}
The conventional intrinsic spin current will be evaluated in two parts. We will first evaluate the topological term
\begin{equation}
\ba
J^z_{y,1}&\dps =\frac{1}{2}\sum_{m{\bm k}} \langle u_{m \bm k}| \hat{s}^z |u_{m \bm k}\rangle \langle u_{m \bm k}|\{\hat{v}_y,\hat{S}^E\}|u_{m \bm k}\rangle.
\ea
\end{equation}
In the eigenstate basis the velocity has the form $ \hat{{\bm v}} = - \mathrm{i} [{\bf \hat{\mathcal R}}, \hat{H}_{0}]$, with matrix elements
\begin{equation}
	- \mathrm{i} \langle u_{m \bm k}|[{\bf \hat{\mathcal R}}, \hat{H}_{0}] |u_{n \bm k}\rangle= - \mathrm{i} {\bf \mathcal R}^{\bm k}_{mn} (\varepsilon_{n \bm k} - \varepsilon_{m \bm k}).
\end{equation}
From the interband coherence paper, the inter-band part of the density matrix is
\begin{equation}
	\rho_{E{\bm k}, mn} = \frac{e{\bm E}\cdot{\bf \mathcal R}^{\bm k}_{mn} [f(\varepsilon_{m\bm k}) - f(\varepsilon_{n\bm k})]}{\varepsilon_{m \bm k} - \varepsilon_{n\bm k}},
\end{equation}
$J^i_{j,1}$ This can be simplified as
\begin{equation}
	\arraycolsep 0.3ex
	\begin{array}{rl}
	\displaystyle J^i_{j, 1} = & \displaystyle \frac{1}{2} \sum_{mn \bm k} s^i_{mm} \, \big\{ \langle u_{m\bm k}|\hat{v}_j |u_{n\bm k}\rangle \langle u_{n\bm k} |\hat{\rho}_E|u_{m\bm k}\rangle  +    \langle u_{m\bm k} |\hat{\rho}_E|u_{n\bm k}\rangle  \langle u_{n\bm k}|\hat{v}_j |u_{m\bm k}\rangle   \big\} \\ [3ex]
\displaystyle = & \displaystyle -\frac{\mathrm{i}e}{2} \, \sum_{mn\bm k} s^i_{mm} \, \big\{{\bf \mathcal R}^{\bm k}_{mn} {\bm E}\cdot{\bf \mathcal R}^{\bm k}_{nm}[f(\varepsilon_{n\bm k})-f(\varepsilon_{m\bm k})] + {\bm E}\cdot{\bf \mathcal R}^{\bm k}_{mn}{\mathcal R}^{\bm k}_{nm}[f(\varepsilon_{m\bm k}) - f(\varepsilon_{n\bm k})] \big\} \\ [3ex]
\displaystyle = & \displaystyle \frac{\mathrm{i}e}{2} \,\sum_{mn \bm k} \Big\{s^i_{mm} \, ({\bf \mathcal R}^{\bm k}_{mn}{\bm E}\cdot{\bf \mathcal R}^{\bm k}_{nm}-{\bm E}\cdot{\bf \mathcal R}^{\bm k}_{mn} {\bf \mathcal R}^{\bm k}_{nm})f(\varepsilon_{m\bm k})- s^i_{mm} ({\bf \mathcal R}^{\bm k}_{mn} {\bm E}\cdot{\bf \mathcal R}^{\bm k}_{nm}- {\bm E}\cdot{\bf \mathcal R}^{\bm k}_{mn}{\bf \mathcal R}^{\bm k}_{nm})f(\varepsilon_{n\bm k}) \Big\}\\ [3ex]	 
\displaystyle = & \displaystyle \frac{\mathrm{i}e}{2} \, \sum_{mn \bm k}  \Big\{s^i_{mm} ({\bf \mathcal R}^{\bm k}_{mn}{\bm E}\cdot{\bf \mathcal R}^{\bm k}_{nm}-{\bm E}\cdot{\bf \mathcal R}^{\bm k}_{mn}{\bf \mathcal R}^{\bm k}_{nm})f(\varepsilon_{m\bm k})-s^i_{nn} \, ({\bf \mathcal R}^{\bm k}_{nm}{\bm E}\cdot{\bf \mathcal R}^{\bm k}_{mn}-{\bm E}\cdot{\bf \mathcal R}^{\bm k}_{nm}{\bf \mathcal R}^{\bm k}_{mn})f(\varepsilon_{m\bm k})\Big\}.
\end{array}
\end{equation}
Now take the electric field to be along $x$ and the current to be along $y$
\begin{equation}
	\arraycolsep 0.3ex
	\begin{array}{rl}
		\displaystyle J^z_{y, 1} = & \displaystyle \frac{\mathrm{i}eE_x}{2} \, \sum_{mn \bm k} \Big\{ s^z_{mm} ({\mathcal R}^y_{mn} {\mathcal R}^x_{nm}- {\mathcal R}^x_{mn} {\mathcal R}^y_{nm}) f(\varepsilon_{m\bm k})  
		+({\mathcal R}^y_{mn} s^z_{nn} {\mathcal R}^x_{nm}- {\mathcal R}^x_{mn}s^z_{nn}{\mathcal R}^y_{nm}) f(\varepsilon_{m\bm k})\Big\}.
	\end{array}
\end{equation}
Consider the product of two Berry connections in the first term ${\mathcal R}^y_{mn} {\mathcal R}^x_{nm}\rightarrow \dbkt{\pd{u_{m\bm k}}{k_y}}{\pd{u_{m\bm k}}{k_x}}$,
\begin{equation}\label{J1}
	\arraycolsep 0.3ex
	\begin{array}{rl}		
		\displaystyle J^z_{y, 1} = & \displaystyle \frac{\mathrm{i}eE_x}{2} \, \sum_{mn \bm k}  s^z_{mm} \Big[\dbkt{\pd{u_{m\bm k}}{k_y}}{\pd{u_{m\bm k}}{k_x}}  - \dbkt{\pd{u_{m\bm k}}{k_x}}{\pd{u_{m\bm k}}{k_y}} \Big] f(\varepsilon_{m\bm k}) 
		+ \Big[{\mathcal R}^y_{mn} s^z_{nn}{\mathcal R}^x_{nm}- {\mathcal R}^x_{mn}s^z_{nn}{\mathcal R}^y_{nm}\Big] f(\varepsilon_{m\bm k}) \\ [3ex]
		= & \displaystyle - \frac{eE_x}{2} \, \sum_{m\bm k} \Big[s^z_{mm}\Omega^z_{m \bm k}+ \Sigma^{z}_{m \bm k}\Big] f(\varepsilon_{m\bm k}).		
	\end{array}
\end{equation}
We have defined $\Sigma^{z,z}_{m\bm k}=\sum_{{\bm k},n} \mathrm{i}[\mathcal{R}^x_{mn}s^z_{nn}\mathcal{R}^y_{nm}-\mathcal{R}^y_{mn}s^z_{nn}\mathcal{R}^x_{nm}]$.
Then the other term of the conventional intrinsic spin current has the form
\begin{equation}\label{J2}
\ba
&\dps J^z_{y,2} =\frac{1}{2}\sum_{\bm k}\sum_{m\neq n}\langle u_{m\bm k}|\hat{s}^z|u_{n\bm k}\rangle \langle u_{n\bm k}|\{\hat{v}_y,\hat{\rho}_E|u_{m\bm k}\rangle\\[3ex]
&\dps =\frac{1}{2}\sum_{\bm k}\sum_{m\neq n}s^z_{mn}[v^y_{nn}+ v^y_{mm}]\rho_{E{\bm k},nm} +\frac{1}{2}\sum_{\bm k}\sum_{m'\neq m,n}s^z_{mn}[v^y_{nm'}\rho_{E{\bm k},m'm}+\rho_{E{\bm k},nm'}v^y_{m'm}].
\ea
\end{equation}
\subsection{Torque dipole correction 1: ${\bm Q}$-dependent part of the non-equilibrium density matrix}
The torque dipole term is $I^i_j =\frac{1}{2} {\rm Tr} \hat{\rho} \{ \hat{t}^i, \hat{r}_j\}$. The density matrix $\rho_{E{\bm k}_+{\bm k}_-}$ to first order in ${\bm Q}$ can be simplified as
\begin{equation}
\ba
\langle m|\rho_{E{\bm k}_+{\bm k}_-}|n\rangle&\dps =\frac{-\mathrm{i}\hbar}{\varepsilon_{m\bm k}-\varepsilon_{n\bm k}}\Big\{-\frac{\mathrm{i}{\bm Q}}{2}\cdot\Big\{v_{\bm k} ,\rho_{E\bm k}\Big\}_{mn}\Big\},
\ea
\end{equation}
the velocity operator can be separated into a part the diagonal in the band index and a part off-diagonal velocity in the band index. The derivative of $\rho_{E{\bm k}_+{\bm k}_-}$ with respect to $\bm Q$ is expressed as
\begin{equation}
\ba
\langle m|\pd{\rho_{{E\bm k}_+{\bm k}_-}}{\bm Q}|n\rangle
&\dps =-\hbar\frac{[v^j_{mm}+v^j_{nn}]\rho_{E{\bm k},mn}}{2(\varepsilon_{m\bm k}-\varepsilon_{n\bm k})} -\hbar\sum_{m'\neq mn}\frac{v^j_{mm'}\rho_{E{\bm k},m'n} + 
\rho_{E{\bm k},mm'} v^j_{m'n}}{2(\varepsilon_{m\bm k}-\varepsilon_{n\bm k})}.
\ea
\end{equation}
By feeding 
$t^i_{nm}=\frac{\mathrm{i}}{\hbar}\langle u_{n\bm k}|[\hat{H}_0,\hat{s}^i]|u_{m\bm k}\rangle=\frac{\mathrm{i}}{\hbar}(\varepsilon_{n\bm k}-\varepsilon_{m\bm k})s^i_{nm}$ into the trace,
\begin{equation}
\ba
 I^i_j&\dps= \mathrm{i}\sum_{mn \bm k}\frac{\mathrm{i}}{\hbar}(\varepsilon_{m\bm k}-\varepsilon_{n\bm k})s^i_{mn}\langle u_{n\bm k}|\pd{\rho_{E{\bm k}_+{\bm k}_-}}{\bm Q}|u_{m \bm k}\rangle\\[3ex]
 &\dps =-\frac{1}{2}\sum_{mn \bm k}s^i_{mn} [v^j_{mm}+v^j_{nn}]\rho_{E{\bm k},nm}-\frac{1}{2}\sum_{\bm k}\sum_{m\neq n}\sum_{m'\neq mn}s^i_{mn}\Big[v^j_{nm'}\rho_{E{\bm k},m'm} + \rho_{E{\bm k},nm'} v^j_{m'm}\Big].
 \ea
\end{equation}
If the electric field is along ${\bm x}$ direction, we have 
\begin{equation}\label{I1}
\ba
 I^z_y&\dps=-\frac{1}{2}\sum_{mn \bm k}s^z_{mn} [v^y_{mm}+v^y_{nn}]\rho_{E{\bm k},nm}-\frac{1}{2}\sum_{\bm k}\sum_{m\neq n}\sum_{m'\neq mn}s^z_{mn}\Big[v^y_{nm'}\rho_{E{\bm k},m'm} + \rho_{E{\bm k},nm'} v^y_{m'm}\Big].
\ea
\end{equation}
This will exactly cancel the intrinsic torque contribution (\ref{J2}). The only spin current term that is left is the topological term (\ref{J1}) and written in general vector form
\begin{equation}
	\bm{\mathcal J}^i = \frac{e{\bm E}}{2} \times \sum_{m{\bm k}} \Big[s^i_{mm} \bm{\Omega}_{m\bm k}+ \bm{\Sigma}^{i}_{m\bm k}\Big]f(\varepsilon_{m\bm k}).
\end{equation}
\subsection{Torque dipole correction 2: $\bm Q$-dependent part of the equilibrium density matrix}
The equilibrium density matrix will have a correction that is linear in the infinitesimal wavevector $\bm Q$ from the driving term $- \frac{\mathrm{i}{\bm Q}}{2\hbar}\cdot\bigg\{\frac{DH_{0{\bm k}}}{D{\bm k}}, \rho_{0{\bm k}} \bigg\}$
\begin{equation}
    \rho^{mn}_{\bm Q}= - \mathrm{i}\frac{\bm Q}{2}\cdot\mathcal{R}^{\bm k}_{mn}\left[f(\varepsilon_{m\bm k}) +f(\varepsilon_{n\bm k})\right]\,.
\end{equation}
Here $\rho^{mn}_{\bm Q}$ is off-diagonal in the band index. This gives an extra driving term in the kinetic equation
\begin{equation}
\pd{\rho_{{\bm k}_+{\bm k}_-}}{t}+\frac{\mathrm{i}}{\hbar}[H_{0{\bm k}}, \rho_{{\bm k}_+{\bm k}_-}] + \frac{\mathrm{i}{\bm Q}}{2\hbar}\cdot\bigg\{\frac{DH_{0{\bm k}}}{D{\bm k}}, \rho_{{\bm k}_+{\bm k}_-} \bigg\}=d_{E{\bm k}}.
\end{equation}
$d_{E{\bm k}}$ will have extra contribution because of linear term in the infinitesimal wavevector $\bm Q$ which is
\begin{equation}
\begin{aligned}
   d^{mm'}_{E{\bm Q}}= - \frac{\mathrm{i}}{\hbar} \left[H_E,\rho_{\bm Q}\right]_{mm^{\prime}}=\frac{e E_i}{\hbar} \frac{D \rho^{mm^{\prime}}_{\bm Q}}{D k_i}\,.
\end{aligned}
\end{equation}
Evaluating the two components of 
the covariant derivative, this driving term gives
\begin{equation}
\begin{aligned}
    \left[\frac{\partial \rho_{Q_j}}{\partial k_i}\right]^{mm^{\prime}} = -\frac{\mathrm{i} Q_j}{2}\left[\frac{\partial\mathcal{R}_{mm^\prime}^j}{\partial k_i}\left[f(\varepsilon_{m\bm k}) + f(\varepsilon_{m'\bm k})\right]+\mathcal{R}_{mm^\prime}^j\left(\frac{\partial f(\varepsilon_{m\bm k})}{\partial k_i} + \frac{\partial f(\varepsilon_{m'\bm k}) }{\partial k_i}\right) \right]\,,
\end{aligned}
\end{equation}
and
\begin{equation}
\begin{aligned}
    -\frac{\mathrm{i}}{\hbar}\left[\mathcal{R}_i,\rho_{Q_j}\right]^{mm^{\prime}} =& -\frac{Q_j}{2\hbar} \sum_{n,m^\prime \neq n} \mathcal{R}^i_{mn}\mathcal{R}^j_{nm^\prime}\big[f(\varepsilon_{n\bm k})+f(\varepsilon_{m'\bm k})\big]+\frac{Q_j}{2\hbar}\sum_{n,m\neq n}\mathcal{R}^j_{mn}\mathcal{R}^i_{nm^\prime}\big[f(\varepsilon_{n\bm k})+f(\varepsilon_{m\bm k})\big]\,.
\end{aligned}
\end{equation}
The solution for another part of density matrix $\rho_{E{\bm k}_+{\bm k}_-}$ to first order in ${\bm Q}$ is:
\begin{equation}
    \rho^{mm'}_{E{\bm Q}} = \frac{\hbar d^{mm'}_{E{\bm Q}}}{\mathrm{i}(\varepsilon_{m\bm k}-\varepsilon_{m'\bm k})}\,.
\end{equation}
 Using this we can find the spin current generated by the ${\bm Q}$-dependent part of the equillibrium density matrix
\begin{equation}\label{Ifin}
\begin{aligned}
	I^l_{i} = &  \mathrm{i} \, {\rm Tr} \, t^l_{{\bm k}} \bigg(\pd{\rho_{E {\bm Q}}}{Q_i}\bigg)_{{\bm Q} \rightarrow 0}\\
    =&-\frac{e E_j}{2 \hbar} \sum_{m\neq m^\prime} s^l_{mm^\prime}\bigg[ \frac{\partial\mathcal{R}_{m^\prime m}^j}{\partial k_i}\left(f(\varepsilon_{m\bm k}) + f(\varepsilon_{m'\bm k})\right)+\mathcal{R}_{m^\prime m}^j\left(\frac{\partial f(\varepsilon_{m\bm k})}{\partial k_i} + \frac{\partial f(\varepsilon_{m'\bm k})}{\partial k_i}\right) \\
    & - \mathrm{i} \sum_{n,m\neq n} \mathcal{R}^i_{m^\prime n}\mathcal{R}^j_{nm}\Big(f(\varepsilon_{n\bm k})+f(\varepsilon_{m\bm k})\Big) +\mathrm{i}\sum_{n,m^\prime\neq n}\mathcal{R}^j_{m^\prime n}\mathcal{R}^i_{nm}\Big(f(\varepsilon_{n\bm k})+f(\varepsilon_{m'\bm k})\Big)\bigg]\\
    =&-\frac{e E_j}{2 \hbar} \sum_{m} \frac{\partial f(\varepsilon_{m\bm k})}{\partial k_i} \{s_\text{od}^l,\mathcal{R}^j\}^{mm}+f(\varepsilon_{m\bm k})\{s_\text{od}^l,\frac{\partial \mathcal{R}^{j}}{\partial k_i}\}^{mm} + \frac{\mathrm{i} e E_j}{2 \hbar} \sum_{mm^\prime n} f(\varepsilon_{n\bm k}) \left(\mathcal{R}^j_{mm^\prime}s^l_{\text{od},m^\prime n}\mathcal{R}^i_{nm} - \mathcal{R}^i_{mm^\prime}s^l_{\text{od},m^\prime n}\mathcal{R}^j_{nm}\right)\\
    &+\frac{\mathrm{i} e E_j}{2 \hbar} \sum_{mm^\prime n}f(\varepsilon_{m\bm k}) \left(s^l_{\text{od},mm^\prime}\mathcal{R}^i_{m^\prime n}\mathcal{R}^j_{nm}-\mathcal{R}^j_{mm^\prime}\mathcal{R}^i_{m^\prime n}s^l_{\text{od},nm}\right)\,.
\end{aligned}
\end{equation}
Here the index $\text{od}$ indicates that we only take band off-diagonal elements.  Also $\mathcal{R}^j$ is purely off-diagonal in these expressions as it comes from $\rho_{\boldsymbol{Q}}$. The first two terms in (\ref{Ifin}) can be simplified
\begin{equation}\label{I2part1}
    \begin{aligned}
        & -\frac{e E_j}{2 \hbar} \sum_{m} \frac{\partial f(\varepsilon_{m\bm k})}{\partial k_i} \{s_\text{od}^l,\mathcal{R}_\text{od}^j\}^{mm}+f(\varepsilon_{m\bm k})\left\{s^l_\text{od},\frac{\partial \mathcal{R}_\text{od}^{j}}{\partial k_i}\right\}^{mm}\\
        =& \frac{e E_j}{2 \hbar} \sum_{m} -\frac{\partial}{\partial k_i} \left(f(\varepsilon_{m\bm k}) \{s_\text{od}^l,\mathcal{R}_\text{od}^j\}^{mm} \right)+f(\varepsilon_{m\bm k})\left\{\frac{\partial s_\text{od}^l}{\partial k_i},\mathcal{R}^j_\text{od} \right\}^{mm}\,.
    \end{aligned}
\end{equation}
Here the first term is a total derivative and will vanish when taking the trace. The second two terms in (\ref{Ifin}) can be expressed as
\begin{equation}
    -\frac{\mathrm{i}eE_j}{2\hbar}\sum_m\left\{[\mathcal{R}^i,s_\text{od}^l],\mathcal{R}_\text{od}^j \right\}^{mm}\,,
\end{equation}
together with (\ref{I2part1}) this means that this spin current correction can be written using a covariant derivative as
\begin{equation}
    I^l_{i} =\frac{e E_j}{2 \hbar} \sum_{m} f(\varepsilon_{m\bm k})\left\{\frac{D s^l_\text{od}}{D k_i},\mathcal{R}^j_\text{od}\right\}^{mm}\,.
\end{equation}
The spin current expression can be further simplified by considering the partial derivative of the spin operator in the eigenstate basis
$  \frac{\partial s^{mm^\prime}_\text{od}}{\partial k_i} = \frac{\partial}{\partial k_i} \langle u_{m\bm k}|\hat{s}^i|u_{m^\prime\bm k}\rangle=\mathrm{i}[\mathcal{R}^i,s]^{mm^\prime}_\text{od}$.  Looking at the spin current again we have
\begin{equation}
\begin{aligned}
    I^l_{i} =\frac{e E_j}{2 \hbar} \sum_{m} f(\varepsilon_{m\bm k})\left\{ \frac{\partial s_\text{od}}{\partial k_i} - \mathrm{i}[\mathcal{R}^i,s^l_\text{od}],\mathcal{R}^j_\text{od}\right\}^{mm}=\frac{e E_j}{2 \hbar} \sum_{m} f(\varepsilon_{m\bm k}) \left\{ \mathrm{i}[\mathcal{R}^i,s^l]_\text{od} - \mathrm{i}[\mathcal{R}^i,s^l_\text{od}],\mathcal{R}^j_\text{od}\right\}^{mm}\,.
\end{aligned}
\end{equation}
The trace only takes the diagonal elements of the anti-commutator here and the Berry connection term $\mathcal{R}^j_\text{od}$ is off-diagonal. This means any band diagonal parts of the left hand side of the anti-commutator will not contribute to the trace. We can rewrite the expression for $I^l_{i}$ as
\begin{equation}\label{I2}
\begin{aligned}
    I^l_{i}=&\frac{e E_j}{2 \hbar} \sum_{m} f(\varepsilon_{m\bm k})\left\{ \mathrm{i}[\mathcal{R}^i,s^l]_\text{od} - \mathrm{i}[\mathcal{R}^i,s^l_\text{od}]_\text{od},\mathcal{R}^j_\text{od}\right\}^{mm}\\
    =&\frac{\mathrm{i} e E_j}{2 \hbar} \sum_{m} f(\varepsilon_{m\bm k}) \left\{[\mathcal{R}_\text{od}^i,s^l_\text{d}]_\text{od},\mathcal{R}^j_\text{od}\right\}^{mm}\\
    =&\frac{e E_j}{2 \hbar} \sum_{m,m^{\prime}} f(\varepsilon_{m\bm k})\left(\mathcal{R}^i_{mm^{\prime}}s^l_{m'm'}\mathcal{R}^j_{m^{\prime}m} -s^l_{mm}\mathcal{R}^i_{mm^{\prime}}\mathcal{R}^j_{m^{\prime}m}+\mathcal{R}^j_{mm^{\prime}}\mathcal{R}^i_{m^{\prime}m}s^l_{mm}-\mathcal{R}^j_{mm^{\prime}}s^l_{m'm'}\mathcal{R}^i_{m^{\prime}m}\right)\\
    =&\frac{e E_j}{2 \hbar} \sum_{m,m^{\prime}} f(\varepsilon_{m\bm k})\left(s^l_{mm}(\mathcal{R}^j_{mm^{\prime}}\mathcal{R}^i_{m^{\prime}m}-\mathcal{R}^i_{mm^{\prime}}\mathcal{R}^j_{m^{\prime}m})-(\mathcal{R}^j_{mm^{\prime}}s^l_{m'm'}\mathcal{R}^i_{m^{\prime}m}-\mathcal{R}^i_{mm^{\prime}}s^l_{m'm'}\mathcal{R}^j_{m^{\prime}m})\right)\\
    =&\frac{e E_j}{2 \hbar} \sum_{m} f(\varepsilon_{m\bm k}) s^l_{mm} \epsilon_{jik}\Omega^k_{m\bm k} - f(\varepsilon_{m\bm k})\epsilon_{jik} \Sigma^{k,l}_{m\bm k}
\end{aligned}
\end{equation}
We drop any potential diagonal terms in the commutator $[\mathcal{R}^i,s^l_\text{od}]$ at the third step due to $\mathcal{R}^j_\text{od}$ being off-diagonal, meaning any diagonal terms will not contribute. When combining (\ref{J1}) and (\ref{I2}) we get the total spin current
\begin{equation}
    \mathcal{J}^l_{i} = -\frac{e E_j}{\hbar} \sum_{m \bm k} f(\varepsilon_{m\bm k})\epsilon_{jik} \Sigma^{k,l}_{m \bm k}.
\end{equation}
Or it can be written as
\begin{equation}
    \boldsymbol{\mathcal{J}}^l =\frac{e \bm E}{\hbar} \times \sum_{m \bm k} f(\varepsilon_{m\bm k})\bm\Sigma^{l}_{m\bm k}.
\end{equation}

\section{Luttinger parameters}

\begin{table}[h]
\centering
\caption{\label{LP}Luttinger parameters used in this work \cite{Roland}}
\begin{tabular}{|m{2cm}|m {2cm}|m{2cm}|m{2cm}|m{2cm}|}
\hline
 & GaAs & InAs & InSb & Ge\\
 \hline
$\gamma_1$ & 6.85  &  20.40 & 37.10 & 13.38\\
\hline
$\gamma_2$ & 2.10  &  8.30 & 16.50 & 5.24\\
\hline
$\gamma_3$ & 2.90  &  9.10 & 17.70 & 4.69\\
\hline
\end{tabular}
\end{table}

\end{appendix}
\end{widetext}
\end{document}